\documentclass[11pt,amsmath,amssymb,nofootinbib,superscriptaddress]{revtex4}

\usepackage{ulem}
\usepackage{amsmath}
\usepackage{dcolumn}
\usepackage{bm}
\usepackage{bbm}
\usepackage[all, knot]{xy}
\xyoption{arc}

\begin{document}

\title{Twisted Hopf symmetries of canonical noncommutative
spacetimes\\ and the no-pure-boost principle}

\author{Giovanni Amelino-Camelia}
\affiliation{
Dipartimento di Fisica \\
 Universit\`a di Roma ``La Sapienza"\\
and Sez.~Roma1 INFN\\
P.le A. Moro 2, 00185 Roma , Italy}
\author{Fabio Briscese}
\affiliation{
Dipartimento di Modelli e Metodi Matematici and GNFM \\
 Universit\`a di Roma ``La Sapienza"\\
Via A. Scarpa 16, 00161 Roma , Italy}
\author{Giulia Gubitosi}
\affiliation{
Dipartimento di Fisica \\
 Universit\`a di Roma ``La Sapienza"\\
and Sez.~Roma1 INFN\\
P.le A. Moro 2, 00185 Roma , Italy}
\author{Antonino~Marcian\`o}
\affiliation{
Dipartimento di Fisica \\
 Universit\`a di Roma ``La Sapienza"\\
and Sez.~Roma1 INFN\\
P.le A. Moro 2, 00185 Roma , Italy}
\author{Pierre Martinetti}
\thanks{Supported by EU Marie Curie fellowship EIF-025947-QGNC}
\affiliation{
Dipartimento di Fisica \\
 Universit\`a di Roma ``La Sapienza"\\
and Sez.~Roma1 INFN\\
P.le A. Moro 2, 00185 Roma , Italy}
\author{Flavio Mercati}
\affiliation{
Dipartimento di Fisica \\
 Universit\`a di Roma ``La Sapienza"\\
and Sez.~Roma1 INFN\\
P.le A. Moro 2, 00185 Roma , Italy}

\begin{abstract}
\begin{center}
{\bf Abstract}
\end{center}
\noindent We study the twisted-Hopf-algebra symmetries of
observer-independent canonical spacetime noncommutativity, for
which the commutators of the spacetime coordinates take the form
$[\hat x^\mu,\hat x^\nu]=i\theta^{\mu\nu}$ with observer-independent (and
coordinate-independent) $\theta^{\mu\nu}$. We find that it is
necessary to introduce nontrivial commutators between
transformation parameters and spacetime coordinates, and that the
form of these commutators implies that all symmetry
transformations must include a translation component. We show that
with our noncommutative transformation parameters the Noether
analysis of the symmetries is straightforward, and we compare our
canonical-noncommutativity results with the structure of the
conserved charges and the ``no-pure-boost" requirement derived in
a previous study of $\kappa$-Minkowski noncommutativity.
We also verify that,
while at
intermediate stages of the analysis
we do find terms that depend on the ordering convention adopted in setting
up the Weyl map,
the final result for the conserved charges is reassuringly independent of the choice
of Weyl map and (the corresponding choice of) star product.

\end{abstract}

\maketitle

\newpage
\baselineskip12pt plus .5pt minus .5pt \pagenumbering{arabic} %
\pagestyle{plain}

\section{Introduction}
Over these past few years there has been strong interest in the
study of theories formulated in noncommutative
versions of the Minkowski spacetime. The most studied possibility is
the one of spacetime noncommutativity
of ``canonical"~\cite{wessDefinizio} form
\begin{equation}
[\hat{x}^\mu,\hat{x}^\nu]=i\theta^{\mu\nu}
~,\label{commutationrelation}
\end{equation}
where $\hat{x}^\mu$ are the spacetime coordinates ($\mu \in
\{0,1,2,3\}$, time coordinate $\hat{x}^0$) and $\theta^{\mu\nu}$
is coordinate-independent. The literature on this possibility is
extremely large, also because the same formula
(\ref{commutationrelation}) can actually represent rather
different physical scenarios, depending on the properties
attributed to $\theta^{\mu\nu}$. The earliest studies we are aware
of were actually the ones~\cite{doplich1994} in which richer
properties were attributed to $\theta^{\mu\nu}$, including some
constraints~\cite{doplich1994} on the admissible forms of
$\theta^{\mu\nu}$ and the possibility of nontrivial algebraic
properties~\cite{doplichRecent}. And this picture can find
valuable motivation in the outcome of certain heuristic analyses
of limitations on the localization of a spacetime point in the
quantum-gravity realm~\cite{doplich1994}. The formalism is much
simpler if one analyzes (\ref{commutationrelation}) assuming that
$\theta^{\mu\nu}$ is a (dimensionful) number-valued
tensor~\cite{szabo,douglas,susskind}, and this gives rise to a
picture which could be rather valuable, since it is believed to
provide an accurate effective-theory description of String Theory
in presence of a certain tensor
background~\cite{szabo,douglas,susskind}. The tensor background
breaks the spacetime symmetries in just the way codified by the
tensor $\theta^{\mu\nu}$: the laws of physics are different in
different frames because $\theta^{\mu\nu}$ (transforming like a
Lorentz-Poincar\'e tensor) takes different values in different
frames. The third possibility is for  $\theta^{\mu\nu}$ to be a
number-valued observer-independent matrix. This would of course
require the laws of transformation between inertial observers to
be modified in $\theta$-dependent
manner~\cite{dsr1dsr2,dsrlee,dsrnature}. Preliminary
results~\cite{chaichaTwist,wessTwist,balaTwist} suggest that this
might be accomplished by introducing a description of
translations, boosts and space-rotation transformations based on
the formalism of Hopf algebras.

We intend to focus here on this third possibility, looking for a
deeper understanding of the structure of the Hopf-algebra symmetry
transformations and hoping to set the stage for a more physical
characterization of this novel concept.
In particular, we are interested in establishing similarities
and differences between the Hopf-algebra symmetries of canonical spacetimes
and the Hopf-algebra symmetries
of the so-called $\kappa$-Minkowski spacetime,
for which some of us recently reported
a Noether analysis~\cite{k-Noether,nopure}.

The key ingredient which allowed~\cite{k-Noether,nopure}
the completion (after more than a decade
of failed attempts) of some Noether analyses
in the $\kappa$-Minkowski
case is the introduction of ``noncommutative
transformation parameters" with appropriate nontrivial
commutators with the spacetime coordinates.
And interestingly
the form of the commutators between transformation parameters and
spacetime coordinates turns out to be incompatible with the possibility of a
pure boost.

We intend to show here that analogous structures appear in the
analysis of the Hopf-algebra symmetries of observer-independent
canonical  spacetime noncommutativity. In this case we find that
neither a pure boost nor a pure rotation are allowed, and,
combining these results with the ones previously obtained for
$\kappa$-Minkowski, we conjecture a (limited) universality of a
no-pure-boost uncertainty principle for Hopf-algebra symmetries of
noncommutative Minkowski-like spacetimes.

We also stress the significance of the fact that our Noether analysis
derives  10 conserved charges from the Poincar\'e like Hopf-algebra symmetries.
This provides encouragement for the idea that these Hopf-algebra symmetries
are truly meaningful in characterizing observable aspects of the relevant
theories, contrary to what feared by some authors (see, {\it e.g.}, Ref.~\cite{kosiNOsymm}),
who had argued that the Hopf-algebra structures
encountered in the study of canonical noncommutative spacetimes
might be just a fancy mathematical formalization of
a rather trivial break down of symmetry.

Guided again by intuition developed in our previous studies
of $\kappa$-Minkowski~\cite{k-Noether,nopure,aadluna}
we also expose an ordering issue for the so-called
classical-action description of the generators of symmetry transformations
in canonical noncommutative spacetime.
While this issue should be carefully monitored in future analyses
of other aspects of theories in canonical noncommutative spacetimes,
we reassuringly find that our result for the charges has no dependence
on this choice of ordering prescription.

\section{Twisted-Hopf symmetry algebra and ordering issues}
Our first task is to show that the much
studied~\cite{chaichaTwist,wessTwist,balaTwist}
``twisted'' Hopf algebra of (candidate) symmetries of canonical noncommutative
spacetime can be obtained by introducing rules
of ``classical action''~\cite{aadluna} for the generators of
the symmetry algebra.
We start by observing that
the fields one considers in constructing theories in a
canonical noncommutative spacetime
can be written in the form~\cite{wessDefinizio}:
\begin{equation}
\Phi(\hat{x}) = \int d^4 k \,\tilde \Phi_w(k) e^{i k \hat x}
\label{fourierW}
\end{equation}
by introducing ordinary (commutative) ``Fourier parameters" $k_\mu$ \footnote{We
use $\hat x$
for noncommuting coordinates, $x$ for the auxiliary commuting ones.}.

This associates to any given function $\Phi(\hat{x})$
a ``Fourier transform" $\tilde{\Phi}_w(k)$, and it is customary to take this one step
further by using this as the basis for an association, codified in a ``Weyl map" $\Omega_w$,
 between the noncommutative functions $\Phi(\hat{x})$
of interest and some auxiliary commutative functions $\Phi^{(comm)}_w(x)$:
\begin{equation}
\Phi(\hat{x}) = \Omega_w \left(\Phi^{(comm)}_w(x)\right) \equiv
\Omega_w \left( \int d^4 k\, \tilde\Phi_w (k) e^{ikx}
\right)=\int d^4k \,\tilde\Phi_w (k) e^{ik\hat x}
\label{weylmap}
\end{equation}
It is easy to verify that
this definition of the Weyl map $\Omega_w$ acts on a given commutative function
by giving a noncommutative function with full symmetrization
(``Weyl ordering")
on the noncommutative
spacetime coordinates ({\it e.g.},  $\Omega_w(e^{ikx})=e^{i k\hat x}$
and $\Omega_w (x_1 x_2^2)=\frac{1}{3}\left(\hat x_2^2 \hat
x_1+\hat x_2\hat x_1 \hat x_2 + \hat x_1 \hat x_2^2\right)$).

We shall stress that it is also legitimate to consider Weyl maps with other
ordering prescriptions, but before we do that let us first use $\Omega_w$
for  our description of the relevant  twisted Hopf algebra.
This comes about by introducing rules of  ``classical action" for the generators of
translations and space rotations
and boosts:\footnote{In light of (\ref{fourierW}) one obtains a fully general rule of action
of operators by specifying their action only on the exponentials $e^{ik\hat x}$.
Also note that we adopt a standard compact notation for antisymmetrized
indices: $A_{[\alpha \beta]}\equiv A_{\alpha \beta}-A_{ \beta \alpha}$.}:
\begin{eqnarray}
&P_\mu^{(w)} e^{ik\hat x}&\equiv P_\mu^{(w)} \Omega_w (e^{ikx})\equiv \Omega_w(P_\mu
e^{ikx})=\Omega_w(i\partial_\mu e^{ikx}) \label{classicalWa} \\
&M_{\mu\nu}^{(w)}e^{ik\hat x}&\equiv M_{\mu\nu}^{(w)}\Omega_w
(e^{ikx})\equiv \Omega_w(M_{\mu\nu}
e^{ikx})=\Omega_w(ix_{[\mu}\partial_{\nu]} e^{ikx}) ~.
\label{classicalWb}
\end{eqnarray}
Here the antisymmetric ``Lorentz-sector" matrix of operators $M^{(w)}_{\mu\nu}$
is composed as usual by the space-rotation
generators $R_{i}^{(w)}=\frac{1}{2}\epsilon_{ijk}M_{j k}^{(w)}$
and the boost generators $N_{i}^{(w)}=M_{0 i}^{(w)}$. The rules of action codified in
(\ref{classicalWa})-(\ref{classicalWb}) are said to be ``classical
actions according to the Weyl map $\Omega_w$" since they indeed
reproduce the corresponding classical rules of action within the
Weyl map.

It is easy to verify that the generators introduced
in (\ref{classicalWa})-(\ref{classicalWb})
satisfy the same commutation relations
of the classical Poincar\'e algebra:
\begin{eqnarray}
\left[ P^{(w)}_\mu , P^{(w)}_\nu \right]&=&0 \nonumber\\
\left[ P^{(w)}_\alpha,M^{(w)}_{\mu\nu}  \right]&=& i \eta_{\alpha [\mu} P^{(w)}_{\nu]} \nonumber \\
\left[M^{(w)}_{\mu\nu},M^{(w)}_{\alpha \beta}   \right]&=& i \left( \eta_{\alpha [\nu}M^{(w)}_{\mu]\beta}
+ \eta_{\beta [\mu}M^{(w)}_{\nu] \alpha}  \right) ~.
\end{eqnarray}
However, the
action of Lorentz-sector generators does not comply with Leibniz
rule,
\begin{eqnarray}
M_{\mu \nu}^{(w)}\left(e^{ik\hat x}e^{i q \hat x}\right)&=& \left(M_{\mu \nu}^{(w)}
e^{ik\hat x}\right)e^{iq\hat x}+  e^{ik\hat x} \left(M_{\mu \nu}^{(w)} e^{iq\hat x}\right)+\nonumber\\
&&-\frac{1}{2}\theta^{\alpha\beta}\left[ \eta_{\alpha [\mu}\left(P_{\nu]}^{(w)}
e^{ik\hat x}\right)\left( P_\beta^{(w)} e^{iq\hat x} \right) +\left(P_\alpha^{(w)}
e^{ik\hat x}\right) \eta_{\beta [\mu} \left(P_{\nu]}^{(w)}
e^{iq\hat x}\right)\right] ~,\label{actiononexponentialsM}
\end{eqnarray}
as one easily verifies using the fact that from
(\ref{commutationrelation}) it follows that
\begin{equation}
e^{ik\hat x}e^{iq\hat x}=e^{i(k+q)\hat x}
e^{-\frac{i}{2}k^\mu \theta_{\mu\nu}q^\nu}\equiv \Omega_w(e^{i(k+q) x}
e^{-\frac{i}{2}k^\mu \theta_{\mu\nu}q^\nu}).\label{ProdottoDiEsponenziali}
\end{equation}

For the translation generators instead Leibniz rule is satisfied,
\begin{eqnarray}
P_\mu^{(w)} \left(e^{ik\hat x}e^{i q \hat x}\right)&=&\left(P_\mu^{(w)}  e^{ik\hat x}\right)e^{iq\hat x}
+  e^{ik\hat x}\left( P_\mu^{(w)} e^{iq\hat x}\right) ~,
\label{actiononexponentialsP}
\end{eqnarray}
as one could have expected from the form of the commutators
(\ref{commutationrelation}), which is evidently compatible with
classical translation symmetry (while, for observer-independent
$\theta^{\mu\nu}$, it clearly requires an adaptation of the
Lorentz sector.)

In the relevant literature
observations of the type reported in (\ref{actiononexponentialsM}) and (\ref{actiononexponentialsP})
are often described via
 an Hopf algebraic structure, specifying the coproduct
\begin{eqnarray}
\Delta P_\mu^{(w)}&=&P_\mu^{(w)} \otimes \mathbbm{1}
+  \mathbbm{1} \otimes P_\mu^{(w)}  ~, \nonumber \\
\Delta M_{\mu \nu}^{(w)}&=& M_{\mu \nu}^{(w)} \otimes \mathbbm{1}
+  \mathbbm{1} \otimes M_{\mu \nu}^{(w)}-\frac{1}{2}\theta^{\alpha\beta}\left[ \eta_{\alpha [\mu}
P_{\nu]}^{(w)}\otimes P_\beta^{(w)} +P_\alpha^{(w)}\otimes \eta_{\beta [\mu}P_{\nu]}^{(w)}\right] ~.
\label{Coproducts}
\end{eqnarray}
Antipode and counit, the other two building blocks needed for a Hopf algebra, can also
be straightforwardly introduced~\cite{giuliatesi}, but do not play a role in
the analysis we are here reporting.

It turns out that the
coproducts (\ref{Coproducts}) are describable as a deformation
of  the classical Poincar\'e Lie algebra  by the following twist element:
\begin{equation}
\mathcal F = e^{\frac{i}{2} \theta^{\mu\nu}P_\mu^{(w)} \otimes P_\nu^{(w)}}.
\label{twistelement}
\end{equation}
The form of the twist element is most easily obtained from the structure of the
star product, which is a way to reproduce the rule of product of noncommutative
functions within the Weyl map: $e^{ik\hat x}e^{iq\hat x}\equiv \Omega_w(e^{ikx} \star e^{iqx})$.
From  (\ref{ProdottoDiEsponenziali}) we see that our star product must be such
that $e^{ikx}\star e^{iqx}=e^{i(k+q)x}e^{-\frac{i}{2}k^\mu\theta_{\mu\nu}q^\nu}$,
and denoting by $\bar{\mathcal F}\equiv \sum(\bar f_1 \otimes \bar f_2) $
the representation of the inverse of the twist element $\mathcal F ^{-1} $
on $\mathcal A\otimes \mathcal A$ (where $\mathcal A$ is the algebra of
commutative functions $f(x)$) we must have~\cite{chaichaTwist}
that $\Omega_w(g(x)*h(x))=\Omega_w(\sum (\bar f_1 (g))(\bar f_2(h)))$,
from which (\ref{twistelement}) follows.

Hopf algebras that are obtained from a given Lie algebra by exclusively acting
with a twist element preserve the form of the commutators among generators,
so that all the structure of the deformation is codified in the coproducts.
And these coproducts are structured in such a way that for
a generator $G_\theta$, obtained twisting $G$,
the coproduct $\Delta_\theta$ has the
form $\Delta_\theta(G_\theta) =\mathcal F \Delta(G) \mathcal F^{-1}$.

Having established that by introducing ``classical action
according to $\Omega_w$" for translations, space-rotations
and boosts one obtains a certain set of generators for a twisted Hopf algebra, it is
natural to ask if something different is encountered
if these generators are introduced with classical
action according to a different Weyl map,
such as the Weyl map  $\Omega_1$ defined
by $\Omega_1(e^{ikx})=e^{ik^A \hat x_A}e^{i k^1 \hat x_1}$,
where $A=0,2,3$.

A given filed $\Phi(\hat x)$ which according to the Weyl map  $\Omega_w$ has
Fourier transform $\tilde \Phi_w (k)$
(in the sense of (\ref{fourierW})), according to $\Omega_1$
has a different Fourier transform $\Phi_1(k)$,
\begin{equation}
 \Phi(\hat x)=\int d^4k \tilde \Phi_1(k) e^{ik^A\hat x_A}e^{ik^1 \hat x_1},
 \label{joc99}
\end{equation}
and, since $e^{ik\hat x}=e^{ik^A\hat x_A} e^{ik^1 \hat x_1}e^{\frac{i}{2} k^A k^1 \theta_{A1}}$,
 the two Fourier transforms are simply related:
\begin{equation}
 \tilde \Phi_1(k)=\tilde \Phi_w (k)e^{-\frac{i}{2}k^A k^1 \theta_{A1}} ~.
\end{equation}

Denoting by $P^{(1)}_\mu$ and $M^{(1)}_{\mu \nu}$ the generators with ``classical action
according to $\Omega_1$" one easily finds that they also leave invariant
the commutation relations (\ref{commutationrelation}).
And, as most easily verified~\cite{giuliatesi}
through a simple analysis of the action of these generators
on $e^{ik\hat x}=\Omega_w(e^{ikx})=\Omega_1(e^{ikx}e^{\frac{i}{2}k^Ak^1\theta_{A1}})$,
the following relations hold:
\begin{eqnarray}
P^{(1)}_\mu&=&P^{(w)}_\mu \nonumber\equiv P_\mu ~,\\
M^{(1)}_{\mu \nu}&=& M^{(w)}_{\mu \nu}
+ \frac{1}{2}\theta^{A1}[\eta_{1[\mu} P_{\nu]} P_A+\eta_{A[\mu} P_{\nu]} P_1 ] ~.
\end{eqnarray}

Setting aside the difference between $M^{(1)}_{\mu \nu}$ and $M^{(w)}_{\mu \nu}$,
one could say that the construction based on the two
Weyl maps $\Omega_w$ and  $\Omega_1$ lead to completely analogous structures.
Again one easily uncovers the structure of a twisted Hopf algebra,
the commutators of generators are undeformed,
and all the structure of the deformation is in a coproduct relation,
which in the case of the $\Omega_1$ map takes the form
\begin{eqnarray}
\Delta M^{(1)}_{\mu \nu}&=& M^{(1)}_{\mu \nu} \otimes \mathbbm{1}
+  \mathbbm{1} \otimes M^{(1)}_{\mu \nu}
-\frac{1}{2}\theta^{\alpha\beta}\left[ \eta_{\alpha [\mu}P_{\nu]}\otimes P_\beta
+P_\alpha\otimes \eta_{\beta [\mu}P_{\nu]}\right]+\nonumber \\
&+&\frac{1}{2}\theta^{A1}\left[    \eta_{A [\mu} P_{\nu]}\otimes P_1
+ \eta_{1 [\mu}P_{\nu]}\otimes P_{A}+  P_1 \otimes P_{[\nu} \eta_{\mu]A}
+ P_A \otimes P_{[\nu} \eta_{\mu] 1}\right] ~.
\end{eqnarray}
This may be viewed again as the result of ``twisting", which in this case would
be due to the following twist element
\begin{equation}
\mathcal F_1 =e^{\frac{i}{2}\theta_{AB}P^A\otimes P^B}e^{-i\theta_{A1}P^1\otimes P^A},
\end{equation}
where $A,B={0,2,3}$.

The two sets of generators
$\{P_\mu , M^{(1)}_{\mu \nu} \}$ and $\{P_\mu , M^{(w)}_{\mu \nu} \}$
can be meaningfully described as two bases of generators for the same twisted Hopf algebra.
However, we shall keep track of the structures we encounter
as a result of the difference between $M^{(1)}_{\mu \nu}$ and $M^{(w)}_{\mu \nu}$,
which, since these differences merely amount to a choice of ordering convention,
we expect not to affect the observable features of our theory.

\section{Noncommutative transformation parameters}
Our analysis of canonical noncommutativity will be guided
by the description of symmetry transformations for $\kappa$-Minkowski
spacetime noncommutativity reported by some of us in Refs.~\cite{k-Noether,nopure}.
After the failures of several other attempts,
the criteria adopted in Refs.~\cite{k-Noether,nopure} finally allowed
us to complete successfully the Noether analysis, including the identification
of some conserved (time-independent) charges associated to the symmetries.
We shall therefore assume that those criteria should be also adopted in
the case of canonical noncommutativity.

In Refs.~\cite{k-Noether,nopure} $\kappa$-Poincar\'e symmetry
transformations
of a function $f({\hat{x}})$ of the $\kappa$-Minkowski
spacetime coordinates were parametrized as follows:
\begin{equation}
df(\hat x)=i\left(\gamma^\mu P_\mu+\sigma_j R_j+\tau_k N_k \right)f(\hat x) , \label{differenzialeP}
\end{equation}
 where $\gamma_\mu$, $\sigma_j$, $\tau_k$ are the transformation parameters (respectively translation,
 space-rotation and boost  parameters), and $P_\mu$, $R_j$, $N_k$ are, respectively, translation,
 space-rotation and boost generators.

The properties of the
 transformation parameters $\gamma_\mu$, $\sigma_j$ and $\tau_k$ were derived~\cite{k-Noether,nopure} by
imposing Leibniz rule on the $d$,
\begin{equation}
d(f(\hat x)g(\hat x))=(df(\hat x))g(\hat x)+f(\hat x)(dg(\hat x)) ~.
\label{leibniz}
\end{equation}
It turned out that this requirement cannot be satisfied by
standard (commutative) transformation parameters, so Refs.~\cite{k-Noether,nopure}
introduced the concept of ``noncommutative transformation
parameters" as the most conservative generalization of the standard
concept of transformation
parameters that would allow to satisfy the Leibniz rule.
These noncommutative transformation parameters were required to still
act only by (associative)
multiplication on the spacetime coordinates, but were allowed
to be subject to nontrivial rules of commutation with the spacetime coordinates.
An intriguing aspect of the commutators between
transformation parameters and spacetime coordinates derived in Refs.~\cite{k-Noether,nopure}
is that they turn out to be incompatible with the possibility of a pure boost.
The structure of $\kappa$-Minkowski spacetime does allow pure translations
and pure space-rotations, but
when the
boost parameters are not set to zero then also the space-rotation
parameters must not all be zero.

We intend to introduce here an analogous description of
the twisted Hopf symmetry transformations
of canonical spacetimes.
We consider first the case of the generators $P_\mu,M_{\mu\nu}^{(w)}$,
with classical action according to the Weyl map $\Omega_w$, and we start by
analyzing the case of a pure translation transformation:
\begin{equation}
d_Pf(\hat x)=i\gamma^\mu_{(w)} P_\mu f(\hat x).
\end{equation}
Imposing Leibniz rule, because of
 the triviality of the coproduct
of the translation generators (see previous section),
for this case of a pure translation transformation
one easily verifies that the condition imposed by compliance
with  Leibniz rule,
\begin{equation}
[f(\hat x),\gamma^\mu_{(w)}]P_\mu g(\hat x)=0
\end{equation}
is also trivial and is satisfied by ordinary commutative transformation parameters.

For the case of a pure Lorentz-sector transformation,
\begin{equation}
d_Lf(\hat x)=i\omega^{\mu\nu}_{(w)}M_{\mu\nu}^{(w)}f(\hat x),
\end{equation}
by imposing Leibniz rule one arrives at the following nontrivial requirement:
\begin{equation}
[f(\hat x),\omega^{\mu\nu}_{(w)}]M_{\mu\nu}^{(w)}g(\hat x)
=-\frac{1}{2}\omega^{\mu\nu}_{(w)}(\theta_{[\mu}\,^{\sigma}\delta_{\nu]}\,^{\rho}
+\theta^{\rho}\,_{[\mu}\delta_{\nu]}\,^{\sigma})(P_\rho f(\hat x))(P_\sigma g(\hat x)) ~.
\label{leibTroubles}
\end{equation}
This does not admit any solution of the type we are allowing for the
transformation parameters.
In fact, in order to be solutions of (\ref{leibTroubles})
the $\omega^{\mu\nu}_{(w)}$ should be operators with highly nontrivial
action on functions of the spacetime coordinates, rather
than being ``noncommutative parameters",
acting by simple (associative) multiplication on the spacetime coordinates.

We conclude that whereas pure translations are allowed
in canonical spacetimes, the possibility of a pure Lorentz-sector transformation
 is excluded.

We find however that, while pure Lorentz-sector
transformations are not allowed, it is possible to combine
Lorentz-sector and translation transformations.
In fact, if we consider a transformation with
\begin{equation}
df(\hat x)=i \left[\gamma^\alpha_{(w)} P_\alpha
+ \omega^{\mu\nu}_{(w)}M_{\mu\nu}^{(w)}\right]f(\hat x) ~,
\label{tuttodf}
\end{equation}
then the Leibniz-rule requirement takes the form
\begin{equation}
\left[  [f(\hat x),\gamma^\alpha_{(w)}]+\frac{1}{2}\omega^{\mu\nu}_{(w)}(\theta_{[\mu}\,^\alpha
 \delta_{\nu]}\,^\rho + \theta^\rho\,_{[\mu}
  \delta_{\nu]}\,^\alpha)\left(P_\rho f(\hat x)\right)
     \right]P_\alpha g(\hat x)+[f(\hat x),\omega^{\mu\nu}_{(w)}]M_{\mu\nu}^{(w)}g(\hat x)=0 ~,
\end{equation}
which amounts (by imposing that the term proportional
to $P_\alpha g(\hat x)$ and the term proportional to $M_{\mu\nu}^{(w)}g(\hat x)$
be separately null) to the following requirements
\begin{eqnarray}
\left[f(\hat x),\gamma^\alpha_{(w)} \right]&=&
-\frac{1}{2}\omega^{\mu\nu}_{(w)}(\theta_{[\mu}\,^\alpha \delta_{\nu]}\,^\rho
+ \theta^\rho\,_{[\mu} \delta_{\nu]}\,^\alpha)P_\rho f(\hat x) \nonumber \\
\left[f(\hat x),\omega^{\mu\nu}_{(w)}\right]&=& 0 ~. \label{CommParamF}
\end{eqnarray}
And these requirements imply the following properties of the transformation
parameters
\begin{eqnarray}
\left[\hat x^\beta,\gamma^\alpha_{(w)} \right]&
=&-\frac{i}{2}\omega^{\mu\nu}_{(w)}(\theta_{[\mu}\,^\alpha \delta_{\nu]}\,^\beta
+ \theta^\beta\,_{[\mu} \delta_{\nu]}\,^\alpha) \label{nopureLorA} \\
\left[\hat x^\beta,\omega^{\mu\nu}_{(w)}\right]&=& 0 ~,
\label{nopureLorB}
\end{eqnarray}
which are consistent with our criterion for noncommutative transformation
parameters, since they introduce indeed a noncommutativity between transformation
parameters and spacetime coordinates, but in a way that is compatible with
our requirement that the transformation parameters
act only by (associative) multiplication on the spacetime coordinates.

We conclude that Lorentz-sector transformations are allowed but only
in combination with translation transformations.
Indeed (\ref{nopureLorA}) is such that whenever $\omega_{(w)} \neq 0$
then also $\gamma_{(w)}\neq 0$.
 And interestingly
the translation-transformation parameters, which can be commutative in the
case of a pure translation transformation, must
comply with (\ref{nopureLorA}), and therefore be noncommutative parameters,
in the general case of a transformation that combines a translation component
and a Lorentz-sector component.

Since in the
preceding section we raised the issue of possible alternatives to
the $P_\mu,M^{(w)}_{\mu \nu}$ basis,
such as the basis $P_\mu,M^{(1)}_{\mu \nu}$ obtained by a different
ordering prescription in the Weyl map used to introduce the ``classical action"
of the generators, we should stress here that the analysis of transformation
parameters proceeds in exactly the same way
if one works with the basis $P_\mu,M^{(1)}_{\mu \nu}$;
however, the noncommutativity properties of the transformation parameters
are somewhat different.
In the case $P_\mu,M^{(1)}_{\mu \nu}$
one ends up  considering transformations of the form
\begin{equation}
d^{(1)}f(\hat x)=i \left[\gamma^\alpha_{(1)} P_\alpha
+ \omega^{\mu\nu}_{(1)}M_{\mu\nu}^{(1)}\right]f(\hat x)~,
\label{tuttodfCASO1}
\end{equation}
and it is easy to verify that the transformation parameters
must satisfy the following noncommutativity requirements:
\begin{eqnarray}
\left[f(\hat x),\gamma^\alpha_{(1)} \right]&
=&-\frac{1}{2}\omega^{\mu\nu}_{(1)}\Upsilon_{\mu\nu}^{\alpha\rho}P_\rho f(\hat x) \nonumber \\
\left[f(\hat x),\omega^{\mu\nu}_{(1)}\right]&
=& 0, \label{CommParamFbase1}
\end{eqnarray}
where
$\Upsilon_{\mu\nu}^{\alpha\rho}=(\theta_{[\mu}\,^\alpha \delta_{\nu]}\,^\rho
+ \theta^\rho\,_{[\mu} \delta_{\nu]}\,^\alpha)
-\theta^{A1}[\eta_{A [\mu}\delta_{\nu]}\,^\rho \delta_1\,^\alpha
+ \eta_{1 [\mu} \delta_{\nu]}\,^\rho \delta_A\,^\alpha + \eta_{A [\mu}\delta_{\nu]}\,^\alpha
 \delta_1\,^\rho + \eta_{1 [\mu} \delta_{\nu]}\,^\alpha \delta_A\,^\rho]$.

We shall show that, even though the differences
between $P_\mu,M^{(w)}_{\mu \nu}$ and $P_\mu,M^{(1)}_{\mu \nu}$
require different forms of the commutators between transformation parameters
and spacetime coordinates,
these two possible choices of convention for the description
of the symmetry Hopf algebra lead to the same conserved charges.

\section{Conserved charges}
We now test our formulation of twisted-Hopf-algebra symmetry transformations
in the context of a Noether analysis of the simplest and most studied
 theory formulated
in canonical noncommutative spacetime: a theory for a massless scalar field $\phi(\hat x)$
governed by the following Klein-Gordon-like equation of motion:
\begin{equation}
\square \phi(\hat x)\equiv P_\mu P^\mu \phi(\hat x)
=0.\label{equationofmotion}
\end{equation}

Consistently with the analysis reported in the previous section,
we want to obtain conserved charges associated to the transformations of the form
\begin{equation}
\delta\phi(\hat x)=-d\phi(\hat x)=-i \left[\gamma^\alpha_{(w)} P_\alpha
 + \omega^{\mu\nu}_{(w)}M_{\mu\nu}^{(w)}\right]\phi(\hat x) ~, \label{VariazPhi}
\end{equation}
where the first equality holds because the field we are considering is a scalar.

We take as starting point for the Noether analysis the action
\begin{equation}
S=\frac{1}{2}\int d^4\hat x \,\phi(\hat x)\square\phi(\hat x),
\label{action}
\end{equation}
which (as one can easily verify~\cite{giuliatesi})
generates the equation of motion (\ref{equationofmotion})
and is invariant
under the transformation (\ref{VariazPhi}):
\begin{eqnarray}
\delta S &=& \frac{1}{2} \int d^4 \hat x \left( \delta \phi (\hat x)  \square \phi (\hat x)
+ \phi(\hat x)  \square \delta \phi (\hat x) - d (\phi(\hat x)\square\phi(\hat x)) \right)
= \nonumber \\ &=&  \frac{1}{2} \int d^4 \hat x   \phi (\hat x) [ \square , \delta] \phi (\hat x) = 0 .
\label{VariazS}
\end{eqnarray}

Of course, the charges are to be obtained for fields solutions of
the equation of motion, and therefore we can use (\ref{equationofmotion})
to rewrite (\ref{VariazS}) in the following way:
\begin{equation}
\delta S = \frac{1}{2} \int d^4 \hat x \, \phi(\hat x) \square \delta \phi(\hat x)
=  \frac{1}{2} \int d^4 \hat x \,P_\mu \left[ \phi(\hat x) P^\mu \delta \phi(\hat x)
-  (P^\mu \phi(\hat x)) \delta \phi(\hat x) \right] .
\end{equation}
Then using the commutation relations of the infinitesimal
parameters obtained in Eq.~(\ref{CommParamF}) one can further rewrite $\delta S$
in the following insightful manner:
\begin{equation}
\delta S = - i \int d^4 \hat x \left( \gamma_\nu^{(w)} P_\mu T^{\mu \nu}
+ \omega^{\rho \sigma}_{(w)} P_\mu J^\mu_{\rho \sigma}    \right),
\end{equation}
with
\begin{eqnarray}
T^{\mu \nu} &=& \frac{1}{2} \left ( \phi(\hat x) P^\mu P^\nu \phi(\hat x)
-  (P^\mu \phi(\hat x)) P^\nu \phi(\hat x)\right)  ,\nonumber \\
J^\mu_{\rho \sigma} &=& \frac{1}{2} \left( \phi(\hat x) P^\mu
M_{\rho \sigma}^{(w)} \phi(\hat x) - ( P^\mu \phi(\hat x) ) M_{\rho \sigma}^{(w)} \phi(\hat x) \right)
+ \nonumber \\
&& -\frac{1}{4} (\theta_{[\rho}\,^\nu \delta_{\sigma]}\,^\lambda
+ \theta^\lambda\,_{[\rho} \delta_{\sigma]}\,^\nu) \left[ (P_\lambda \phi (\hat x))
P^\mu P_\nu \phi (\hat x)
- (P^\mu P_\lambda \phi (\hat x)) P_\nu \phi (\hat x)  \right] .
\end{eqnarray}

It is rather easy to verify that by spatial integration
of the $0$-th components of the ``currents'' $T^{\mu \nu}$ and $J^\mu_{\rho \sigma} $
one obtains time-independent charges. Denoting this charges
with $Q_\mu$,$K_{\rho \sigma}$,
\begin{equation}
Q_\mu  = \int d^3 \hat x \, T^0_\mu , \qquad K_{\rho \sigma} = \int d^3\hat x \, J^0_{\rho \sigma} ~,
\end{equation}
and using the ordering convention (\ref{fourierW}) for the
Fourier expansion of a generic field which is solution of the equation of motion,
\begin{equation}
\phi (\hat x) = \int d^4 k \, \delta(k^2) \tilde\phi_{(w)}(k) e^{i k \hat x} ~,
\end{equation}
upon integration over the spatial coordinates\footnote{Our spatial Dirac deltas are
such that $\int d^3 \hat x e^{ik^i \hat x_i}=\delta^{(3)}(\vec k)$.}
one finds:
\begin{eqnarray}
Q_\mu &=& \frac{1}{2}  \int d^4 k \, d^4 q \, \delta(k^2) \delta(q^2)
 \tilde{\phi}_{(w)}(k) \tilde{\phi}_{(w)}(q)
   \nonumber\\
 && \left( q^0 - k^0 \right) q_\mu \delta^{(3)}(\vec{k}
 + \vec{q}) e^{i (k^0 + q^0) \hat x_0}e^{\frac{i}{2}(k^0+q^0)(k^i+q^i)\theta_{i0}}
 e^{-\frac{i}{2}k^\mu q^\nu \theta_{\mu\nu}} ~,
\end{eqnarray}
\begin{eqnarray}
K_{\rho \sigma}&=& \frac{1}{2}  \int d^4 k \, d^4 q \, \delta(k^2)  \tilde{\phi}_{(w)}(k) \left[
  i q_{[\rho}   \frac{\partial}{\partial q^{\sigma]}} [ \delta(q^2) \tilde{\phi}_{(w)}(q)]
  - \frac{1}{2} \delta(q^2) (\theta_{[\rho}\,^\nu \delta_{\sigma]}\,^\lambda
  + \theta^\lambda\,_{[\rho} \delta_{\sigma]}\,^\nu) k_\lambda q_\nu \tilde{\phi}_{(w)}(q)  \right] \cdot
   \nonumber\\
   &&\qquad \cdot \left( k^0 - q^0 \right) \delta^{(3)}(\vec{k}
   + \vec{q})   e^{i (k^0 + q^0) \hat x_0}e^{\frac{i}{2}(k^0+q^0)(k^i+q^i)\theta_{i0}}
   e^{-\frac{i}{2}k^\mu q^\nu \theta_{\mu\nu}} ~. \label{SymmetricCharges}
\end{eqnarray}
Then integrating in $d^4k$, and observing that in $K_{\rho \sigma}$
the term $- \frac{1}{2} (\theta_{[\rho}\,^\nu \delta_{\sigma]}\,^\lambda
+ \theta^\lambda\,_{[\rho} \delta_{\sigma]}\,^\nu) k_\lambda q_\nu \tilde{\phi}(q)$
gives null contribution, one obtains:
\begin{eqnarray}
 Q_\mu &=& \frac{1}{2}  \int   \frac{d^4 q}{2|\vec q|} \,  \delta(q^2)\tilde{\phi}_{(w)}(q)
 q_\mu  \left\lbrace  \tilde{\phi}_{(w)}(-\vec q,|\vec q|)  \left( q^0 +|\vec q| \right)
 e^{i (q^0-|\vec q|  ) \hat x_0} e^{-\frac{i}{2}( q^0 -|\vec q|) q^i \theta_{0 i}} +\right.\nonumber \\
&&\left. \qquad+ \tilde{\phi}_{(w)}(-\vec q,-|\vec q|)  \left( q^0 -|\vec q| \right)
e^{i (q^0+|\vec q| ) \hat x_0} e^{-\frac{i}{2}( q^0 +|\vec q|) q^i \theta_{0 i}} \right\rbrace  ,
\end{eqnarray}
\begin{eqnarray}
K_{\rho \sigma} &=&\frac{i}{2}  \int \frac{d^4 q}{2|\vec q|} \,  \delta(q^2) \tilde\phi_{(w)}( q)
q_{[\rho} \left\lbrace (q^0+|\vec q|)   \left[\frac{\partial}{\partial
q^{\sigma]}} \tilde\phi_{(w)}(-\vec q,|\vec q|)\right] e^{i(q^0-|\vec q|)\hat x_0}
e^{-\frac{i}{2}(q^0-|\vec q|)q^i \theta_{0i}}+\right. \nonumber \\
&&\qquad \left.+(q^0-|\vec q|)   \left[\frac{\partial}{\partial
q^{\sigma]}} \tilde\phi_{(w)}(-\vec q,-|\vec q|)\right] e^{i(q^0+|\vec q|)\hat x_0}
e^{-\frac{i}{2}(q^0+|\vec q|)q^i \theta_{0i}}\right\rbrace.
\end{eqnarray}
One can then use the fact that $\delta(q^2)$ imposes $q_0=\pm |\vec q|$,
and the presence of factors of the types $( q^0 -|\vec q|) e^{\alpha ( q^0 +|\vec q|)}$
and $( q^0 +|\vec q|) e^{\alpha ( q^0 -|\vec q|)}$ to obtain the following explicitly time-independent
formulas for the charges:
\begin{eqnarray}
 Q_\mu &=& \frac{1}{2}  \int   \frac{d^4 q}{2|\vec q|} \,  \delta(q^2)\tilde{\phi}_{(w)}(q)
 q_\mu  \left\lbrace  \tilde{\phi}_{(w)}(-\vec q,|\vec q|)  \left( q^0 +|\vec q| \right)
 + \tilde{\phi}_{(w)}(-\vec q,-|\vec q|)  \left( q^0 -|\vec q| \right)  \right\rbrace ,
\end{eqnarray}
\begin{eqnarray}
K_{\rho \sigma} &=&\frac{i}{2}  \int \frac{d^4 q}{2|\vec q|} \,  \delta(q^2) \tilde\phi_{(w)}( q)
q_{[\rho} \left\lbrace (q^0+|\vec q|)   \frac{\partial \tilde\phi_{(w)}(-\vec q,|\vec q|)}{\partial
q^{\sigma]}}  +(q^0-|\vec q|)   \frac{\partial\tilde\phi_{(w)}(-\vec q,-|\vec q|)}{\partial q^{\sigma]}}
  \right\rbrace. \label{charge1}
\end{eqnarray}

\section{Ordering-convention independence of the charges}
In light of the ``choice-of-ordering issue"
 we raised in Section II, which in particular
led us to consider the examples of two possible bases of generators,
the $P_\mu,M^{(w)}_{\mu \nu}$ basis
and the $P_\mu,M^{(1)}_{\mu \nu}$  basis,
and especially considering the fact that
in Section III we found that in different bases the noncommutative
transformation parameters should have somewhat different properties (different form
of the commutators with the spacetime coordinates), it is interesting to verify
whether or not the result for the charges obtained in the previous section working
with the $P_\mu,M^{(w)}_{\mu \nu}$ basis is confirmed by a corresponding analysis
based on the $P_\mu,M^{(1)}_{\mu \nu}$  basis.

When adopting the $P_\mu,M^{(1)}_{\mu \nu}$  basis
the symmetry variation of a field is described by
\begin{equation}
 \delta\phi(\hat x)=-d^{(1)} \phi(\hat x)=-i \left[\gamma^\alpha_{(1)} P_\alpha
 + \omega^{\mu\nu}_{(1)}M_{\mu\nu}^{(1)}\right]\phi(\hat x),
\end{equation}
rather than (\ref{VariazPhi}). And going through the same type of steps
discussed in the previous section the analysis of the symmetry variation
of the action (\ref{action}) then leads to~\cite{giuliatesi} the following formulas
for the currents:
\begin{equation}
T^{\mu \nu\,(1)} = \frac{1}{2} \left ( \phi(\hat x)
P^\mu P^\nu \phi(\hat x) -  P^\mu \phi(\hat x) P^\nu \phi(\hat x)\right)  ~,
\end{equation}
\begin{eqnarray}
J^{\mu\;(1)}_{\rho \sigma} &=& \frac{1}{2} \left( \phi(\hat x)
P^\mu M_{\rho \sigma}^{(1)} \phi(\hat x) -  P^\mu \phi(\hat x)
M_{\rho \sigma}^{(1)} \phi(\hat x) \right)+ \nonumber \\
&& -\frac{1}{4} \Upsilon_{\rho \sigma}^{\nu \lambda} \left[ P_\lambda \phi (\hat x)
P^\mu P_\nu \phi (\hat x) - P^\mu P_\lambda \phi (\hat x)
P_\nu \phi (\hat x)  \right] ~,
\end{eqnarray}
where we used again the compact notation $\Upsilon_{\rho \sigma}^{\nu \lambda}$,
introduced in Section III.

The current $T^{\mu\nu\,(1)}$ is manifestly equal to the
current $T^{\mu\nu}$ obtained in the previous section using
the $P_\mu,M^{(w)}_{\mu \nu}$ basis. Therefore the corresponding
charges also coincide:
\begin{equation}
Q_\mu^{(1)}  \equiv \int d^3 \hat x \, \, T^{0 \, (1)}_\mu  = \int d^3 \hat x \,\, T^{0 }_\mu  =Q_\mu.
\end{equation}

The current $J^{\mu\;(1)}_{\rho \sigma}$ does differ from $J^{\mu}_{\rho \sigma}$
of the previous section in two ways: in place of the factor $\Upsilon_{\rho \sigma}^{\nu \lambda}$
of $J^{\mu\;(1)}_{\rho \sigma}$ one finds in $J^{\mu}_{\rho \sigma}$
the factor $\theta_{[\rho}\,^\nu\delta_{\sigma]}\,^\lambda +\theta^\lambda\,_{[\rho}\delta_{\sigma]}\,^\nu$,
and there are two (operator) factors $M_{\mu\nu}^{(1)}$ in places where in $J^{\mu}_{\rho \sigma}$
one of course has $M_{\mu\nu}^{(w)}$. Still, once again the final result for the charges
is unaffected:
\begin{equation}
K_{\rho\sigma}^{(1)}\equiv \int d^3\hat x\; J^{0\,(1)}_{\rho\sigma} =K_{\rho\sigma}.
\end{equation}
This is conveniently verified by following the ordering conventions
of Eq.~(\ref{joc99}) in
writing the generic solution of the equation of motion,
\begin{equation}
 \phi(\hat x)=\int d^4k\;\delta(k^2)\tilde\phi_{(1)}(k)e^{ik^A\hat x_A}e^{ik^1\hat x_1} ~,
\end{equation}
thereby obtaining, after spatial integration, the following formula for $K_{\rho \sigma}^{(1)}$:
\begin{eqnarray}
K_{\rho \sigma}^{(1)}&=& \frac{1}{2}  \int d^4 k \, d^4 q \, \delta(k^2)  \tilde{\phi}_{(1)}(k)
 \left[   i q_{[\rho}   \frac{\partial}{\partial q^{\sigma]}} [ \delta(q^2) \tilde{\phi}_{(1)}(q)]
 - \frac{1}{2} \delta(q^2) \Upsilon_{\rho\sigma}^{\nu\lambda} k_\lambda q_\nu \tilde{\phi}_{(1)}(q)
   \right]\cdot \label{MapOneCharges} \\ &&\qquad \cdot \left( k^0 -
   q^0 \right) \delta^{(3)}(\vec{k} + \vec{q})
   e^{i (k^0 + q^0) \hat  x_0}e^{\frac{i}{2}(k^0+q^0)(k^i+q^i)\theta_{i0}}e^{-\frac{i}{2}k^\mu q^\nu
    \theta_{\mu\nu}} e^{-\frac{i}{2}(k^Ak^1+q^Aq^1)\theta_{A1}} ~. \nonumber
\end{eqnarray}
And, using observations that are completely analogous to some we discussed
in the previous section, one easily manages~\cite{giuliatesi}
to rewrite $K_{\rho \sigma}^{(1)}$ as follows:
\begin{eqnarray}
K_{\rho \sigma}^{(1)}&=& \frac{i}{2}\int   \frac{d^4 q}{2|\vec q|} \,
  \tilde{\phi}_{(1)}(q) \delta(q^2) q_{[\rho}\left\lbrace (q^0
  + |\vec q|)\frac{\partial}{\partial q^{\sigma]}}\left[\tilde\phi_{(1)}(-\vec q,|\vec q|)
  e^{-i\left(q^A\delta_A^jq^1\theta_{j1}+\frac{1}{2}(|\vec q|+q^0 )\theta_{01}\right)}\right]
   \right.\nonumber \\
&&\qquad \left. + (q^0 -|\vec q|)\frac{\partial}{\partial q^{\sigma]}}
 \left[\tilde\phi_{(1)}(-\vec q,-|\vec q|) e^{-i\left(q^A\delta_A^jq^1\theta_{j1}
 +\frac{1}{2}(-|\vec q|+q^0 )\theta_{01}\right)}\right] \right\rbrace
 ~. \label{charge2}
\end{eqnarray}

This formula for $K_{\rho \sigma}^{(1)}$ is easily shown to reproduce
the corresponding formula for $K_{\rho \sigma}$,
using the fact that, as we showed
in Section III, $\tilde \phi_{(1)}(k)=\tilde\phi_{(w)}(k)e^{-\frac{i}{2}k^Ak^1 \theta_{A1}}$.

This result establishes that
the values of the charges
carried by a given noncommutative field can be treated as objective facts,
independent of the choice of ordering prescription adopted in the analysis.
Working with different ordering prescriptions one arrives at different formulas
(for example (\ref{charge1}) and (\ref{charge2}))
expressing the charges as functionals of the Fourier transform of the fields,
but these differences in the formulas are just such to compensate for
the differences between the
Fourier transforms of a given field that are found adopting different ordering conventions,
and therefore the values of the charges carried by a given noncommutative field
can be stated in an ordering-prescription-independent manner.

\section{Closing remarks}
The fact that we managed to derive a full set of 10 conserved
charges from the twisted-Hopf-algebra symmetries that emerge from observer-independent
canonical noncommutativity certainly provides some encouragement for the idea
that these (contrary to some expectations formulated in the recent literature~\cite{kosiNOsymm})
are genuine physical symmetries. And this viewpoint is strengthened by our
result on the ordering-convention independence of the charges.

The characterization of ``noncommutative transformation parameters" introduced by some
of us in Refs.~\cite{k-Noether,nopure}, for the analysis of theories in $\kappa$-Minkowski
noncommutative spacetime, proved to be valuable also in the present study of
canonical noncommutativity. This type of transformation parameters objectively
does the job (without any need of ``further intervention") of allowing to derive
conserved charges, but it requires still some work for what concerns establishing
its physical implications and its realm of applicability. Is this only an appropriate
recipe for deriving conserved charges? Or can we attribute to it all the roles
that transformation parameters have in a classical-spacetime theory? For example:
does the noncommutativity of these parameters imply that the concept of angle of
rotation around a given axis is ``fuzzy" in a canonical spacetime?

The obstruction encountered in the previous analyses~\cite{nopure}
of $\kappa$-Minkowski spacetime for the realization of a pure boost reappeared
here in the analysis of canonical spacetimes (actually accompanied by an additional obstruction
for the realization of a pure space rotation).
Since, to our acknowledge, canonical and $\kappa$-Minkowski spacetimes are the only examples
of noncommutative versions of Minkowski spacetime that one can single out with some reasonable
physical criteria (see, {\it e.g.}, Ref.~\cite{perspectives} and references therein),
the fact that in both cases pure boosts are not allowed could perhaps motivate
the search of an intuitive argument for the emergence of a universal ``no-pure-boost
principle" from the general structure of spacetime noncommutativity.

A lot remains to be done for a proper characterization of the physical/observable
implications of canonical noncommutativity. By which measurements can a theory with
observer-independent canonical noncommutativity be distinguished from a corresponding
classical-spacetime theory? Of course, this issue would be most naturally addressed
in the context of a theory of quantum fields in the noncommutative spacetime, which
we have postponed to future work.
But even within analyses of classical fields in canonical spacetime,
such as the one we here reported,
 a preliminary investigation of ``observability issues"
could be attempted. Correspondingly new measurement-procedure ideas are needed
in order to test the novel possibility of an obstruction for the realization
of a pure Lorentz-sector transformation.
And more work is also needed for a proper operative characterization of the differences
between the charges here obtained for a theory with observer-independent
canonical noncommutativity
and the corresponding charges of a theory in classical spacetime.


\begin{thebibliography}{1}
\bibitem{wessDefinizio}
J. Madore, S. Schraml, P. Schupp and J. Wess,
Eur.~Phys.~J.~C16 (2000) 161,
hep-th/0001203.

\bibitem{doplich1994} S.~Doplicher, K.~Fredenhagen and J.E.~Roberts,
Commun.\ Math.\ Phys.~172 (1995) 187;
Phys.\ Lett.~B331 (1994) 39.

\bibitem{doplichRecent} S.~Doplicher, arXiv:hep-th/0105251.


\bibitem{szabo}
R.~J.~Szabo,
arXiv:hep-th/0109162.

\bibitem{douglas}
M.R.~Douglas and N.A.~Nekrasov,
Rev.~Mod.~Phys.~73 (2001) 977,
arXiv:hep-th/0106048.

\bibitem{susskind}
A.~Matusis, L.~Susskind and N.~Toumbas,
JHEP 0012 (2000) 002,
arXiv:hep-th/0002075.

\bibitem{dsr1dsr2}
G.~Amelino-Camelia,
Int.\ J.\ Mod.\ Phys.~D11 (2002) 35,
arXiv:gr-qc/0012051;
Phys.\ Lett.~B510 (2001) 255,
arXiv:hep-th/0012238.

\bibitem{dsrlee}
J.~Magueijo and L.~Smolin,
Phys.~Rev.~Lett.~88 (2002) 190403,
arXiv:hep-th/0112090.

\bibitem{dsrnature}
G.~Amelino-Camelia,
Nature 418  (2002) 34.

\bibitem{chaichaTwist}
M.~Chaichian, P.~P.~Kulish, K.~Nishijima and A.~Tureanu,
Phys.\ Lett.~B604 (2004) 98,
arXiv:hep-th/0408069.

\bibitem{wessTwist}
  G.~Fiore and J.~Wess,
  Phys.\ Rev.~D75 (2007) 105022,
  arXiv:hep-th/0701078.

\bibitem{balaTwist}
 A.~P.~Balachandran, T.~R.~Govindarajan, G.~Mangano, A.~Pinzul, B.~A.~Qureshi and S.~Vaidya,
  Phys.\ Rev.~D75 (2007) 045009,
  arXiv:hep-th/0608179.

\bibitem{k-Noether}
A.~Agostini, G.~Amelino-Camelia, M.~Arzano, A.~Marciano and R.~A.~Tacchi,
arXiv:hep-th/0607221, Mod.~Phys.~Lett.~A22 (2007) 1779.

\bibitem{nopure}
  G.~Amelino-Camelia, G.~Gubitosi, A.~Marciano, P.~Martinetti and F.~Mercati,
  arXiv:hep-th/0707.1863.

\bibitem{kosiNOsymm} C.~Gonera, P.~Kosinski, P.~Maslanka and S.~Giller,
Phys.~Lett.~B622 (2005) 192.

\bibitem{aadluna}
A.~Agostini, G.~Amelino-Camelia and F.~D'Andrea,
Int.~J.~Mod.~Phys.~A19 (2004) 5187.
  arXiv:hep-th/0306013.

\bibitem{giuliatesi}
G. Gubitosi, Laurea thesis, in preparation (soon to be posted on the arXiv).


\bibitem{perspectives} G.~Amelino-Camelia, arXiv:gr-qc/0309054.

\end{thebibliography}
\end{document}